\documentclass[amsmath,amssymb,aps,pre,twocolumn,showpacs,floatfix]{revtex4}
\usepackage{mathrsfs}
\usepackage{graphicx}
\usepackage{verbatim}
\usepackage{color}
\newcommand{\lf}{\left}
\newcommand{\rg}{\right}

\newcommand{\ti}{\tilde}

\newcommand{\la}{\langle}
\newcommand{\ra}{\rangle}
\newcommand{\fr}{\frac}
\newcommand{\eps}{\epsilon}
\newcommand{\om}{\omega}
\begin{document}
\title{Quantum mechanical calculation of spectral statistics of modified Kepler problem}
\author{Tao Ma, Rostislav Serota}
\affiliation{Department of Physics, University of Cincinnati, Cincinnati, OH 45221}
\date{Dec. 28, 2010}

\begin{abstract}
Discrepancy between periodic orbit theory and numerical calculation of a modified Kepler problem is cleared by a quantum mechanical calculation. The diagonal approximation already gives a good fit for the numerical calculation. A better result yet is gained by considering the coherent interference between the classical periodic orbits and the Balian-Bloch term. This approach produces improved results for the rectangular billiards as well.
\end{abstract}

\maketitle

\section{Introduction}

In a companion paper, \cite{wickramasinghe08} we investigated level correlations in a modified Kepler (MK) problem and discovered two important effects: quantum-Hall-like jumps in the saturation level rigidity as a function of running energy and persistent oscillations of the level number variance as a function of the interval width. Unlike rectangular billiards (RB),  \cite{wickramasinghe05} where superposition of a number of incommensurate harmonics led to a finite lower limit of the variance, in the modified Kepler problem the level variance can achieve near zero values.

In other words, variance can decrease with the increase of the interval width due to the oscillatory behavior and, moreover, at certain values of the interval width the number of levels is nearly identical for each of the members of the statistical ensemble. This is a highly unusual circumstance from a statistical viewpoint and is due to specifics of the strong semiclassical  spectral correlations in classically integrable systems.

Previously, we used a periodic orbit theory  \cite{wickramasinghe08} to investigate level correlations. However, an important class of periodic obits - radial oscillations for a fixed angle - were omitted. Here we utilized a quantum mechanical approach that confirms a corrected periodic orbits approach. Both are in complete agreement now and are in excellent agreement with the numerical simulation.

The key ingredient of our numerical work is the use of \emph {parametric averaging} (for instance, the aspect ratio of equal area rectangles for RB; for MK see discussion in appendix). This allows us to do statistical sampling for a fixed running energy in the semiclassical spectrum and should be contrasted to the traditional technique of spectral averaging in which the key effects discussed in this work are washed out.

\section{Level correlation function and level number variance}
\subsection{average spectral staircase}

The eigen-energy of Modified Kepler problem $E=2 p\om+ l^2$, where $\om = \sqrt{2\beta}$ \cite{wickramasinghe08}.
The average cumulative number of levels is
\begin{equation}
\la\mathscr{N}(E)\ra =
\fr{ E^{3/2}}{3\om}+\fr{1}{2} (E^{1/2}+\fr{E}{2\om} ) ,
\end{equation}
which is the scaling relation we use in the numerical calculation.
But in the theoretical calculation, only the first term is considered.
We define the scaled energy $\eps=\la\mathscr{N}(E)\ra \approx E^{3/2} / 3\om$.

The average spectral staircase of rectangular billiard (RB) is given in Eq. (5) of Ref. \cite{wickramasinghe05}.

\subsection{variance}
\subsubsection{level fluctuation}

The level fluctuation \cite{wickramasinghe05}
\begin{equation}
\delta\rho = \rho - \la\rho\ra = \delta\rho^{(2)} + \delta\rho^{(1)},
\end{equation}
where $\delta\rho^{(1)}$ is the Balian-Bloch term explained in appendix and $\delta\rho^{(2)}$ is given by

\begin{equation}\label{eq:deltaRhoEpsilon}
\begin{split}
&\delta\rho^{(2)}(\eps)= \fr{1}{(3\om\eps)^{1/3}} (\fr{\om}{M_r})^{1/2} \times \\
&\Bigg[ \sum_{M_\theta \leq \fr{M_r}{\gamma^{(cir)}}}
2 \cos \lf(\pi \fr{M_r^2 (3\om\eps)^{2/3}+ M_\theta^2\om^2}{M_r\om}-\fr{\pi}{4} \rg)+ \\
&\hspace{15pt}\sum_{M_r>0}
  \cos\lf(\fr{\pi M_r (3\om\eps)^{2/3}}{\om}-\fr{\pi}{4} \rg) \Bigg] +\\
&\hspace{1pt}\sum_{M_\theta>0} \fr{1}{(3\om\eps)^{1/3}} \fr{1}{\pi M_\theta}\sin(2\pi M_\theta (3\om\eps)^{1/3}),
\end{split}
\end{equation}
where the second term is due to radial periodic orbits and the third term due to circular periodic orbits.
The level density fluctuation equation is the starting point to calculate spectral statistics.
For RB,
\begin{equation}
\begin{split}
&\delta\rho^{(2)}
= (1/4\eps\pi^3)^{1/4} \\
&\sum_{M_1, M_2 = 1}4  \fr{ \cos(2\pi \sqrt{\ti{M_1}^2 + \ti{M_2}^2}
\sqrt{\fr{4\eps}{\pi}}-\fr{\pi}{4}) }{ (\ti{M_1}^2 + \ti{M_2}^2)^{1/4} } + \\
&\hspace{8pt} \sum_{M_1 = 1}2
\fr{\cos(2\pi \ti{M_1} \sqrt{\fr{4\eps}{\pi}}-\fr{\pi}{4})}{\ti{M_1}^{1/2}} + \\
&\hspace{8pt} \sum_{M_2 = 1}2
\fr{\cos(2\pi \ti{M_2} \sqrt{\fr{4\eps}{\pi}}-\fr{\pi}{4})}{\ti{M_2}^{1/2}} .
\end{split}
\end{equation}
where $\ti{M_1} = M_1 \alpha^{1/4}$ and $\ti{M_2} = M_2 \alpha^{-1/4}$.

\subsubsection{level correlation function}

The correlation function of levels is defined as
$K(\eps; E) =\la \delta\rho(\eps - E/2)\delta\rho(\eps + E/2)\ra$,
where the ensemble average $\la \cdot \ra$ is defined as parametric averaging over an ensemble of $\om$.

Using diagonal approximation,
\begin{equation}\label{eq:KEpsilon}
\begin{split}
&K(\eps; E) \approx \\
&\sum_{M_r = 1}\lf(\lf\lfloor \fr{M_r}{\gamma^{(cir)}}\rg\rfloor + \fr{1}{4} \rg)
\fr{2\om}{M_r (3\om\eps)^{2/3}}
\cos \lf(\fr{2\pi M_r E}{(3\om\eps)^{1/3}}\rg),
\end{split}
\end{equation}
where $\gamma^{(cir)} = (\fr{2\beta}{3\eps})^{1/3}$,
which is the same $\gamma^{(cir)}$ in Eq. (61) of Ref. \cite{wickramasinghe08}.

Comparing Eq. \ref{eq:KEpsilon} with Eq. (62) of \cite{wickramasinghe08}, we have an extra term due to radial periodic orbits, which is the single summation term in Eq. 42 of \cite{wickramasinghe05}.

\subsubsection{variance of MK}
The variance is calculated from
$\Sigma(\eps, E)=\int_{\eps-E/2}^{\eps+E/2}
\int_{\eps-E/2}^{\eps+E/2}
K(\eps_1,\eps_2)d\eps_1d\eps_2$.

\begin{equation}\label{eq:variance_MK}
\Sigma (\eps, E)\approx
\sum_{M_r = 1}\lf(\lf\lfloor \fr{M_r}{\gamma^{(cir)}}\rg\rfloor + \fr{1}{4} \rg)
\fr{2\om}{\pi^2 M_r^3} \sin^2
\lf(\fr{\pi M_r E}{(3\om\eps)^{1/3}}\rg).
\end{equation}

Compared with the Eq. 70 of \cite{wickramasinghe08}, there is the extra contribution from radial periodic orbits.

The new variance equation is
\begin{equation}\label{eq:variance_coherent_MK}
\begin{split}
&\Sigma(\eps, E)
\approx
\sum_{M_r = 1}\lf(\lf\lfloor \fr{M_r}{\gamma^{(cir)}}\rg\rfloor + \fr{1}{8} +
\fr{1}{8} \lf(1 + \sqrt\fr{2M_r}{\om} \rg)^2 \rg)\\
&\fr{2\om}{\pi^2 M_r^3} \sin^2
\lf(\fr{\pi M_r E}{(3\om\eps)^{1/3}}\rg) .
\end{split}
\end{equation}

\subsubsection{numerical simulation of MK}

The ensemble averaging is averaging over an ensemble of $\beta$ from a normal distribution centered at $\beta_0 = 3\times 10^6$ and with a width $\beta_0/20$. For RB, the normal distribution is centered at $1$ with a width $0.2$.
\begin{figure}[htp]
\begin{center}
\includegraphics[width=3.0in]{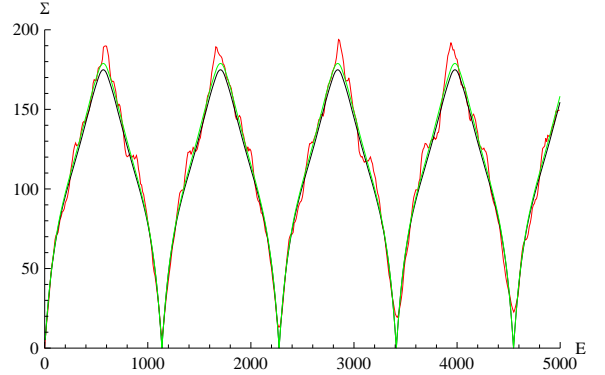}
\end{center}
\caption{We use the same values of $\beta = 3\times 10^6$, and $\eps = 2\times 10^5$ same as in Fig. 2 of Ref. \cite{wickramasinghe08}.
The red line is the numerical result.
The black line the old theoretical result from Eq. \ref{eq:variance_MK}.
The green line is the new theoretical result from Eq. \ref{eq:variance_coherent_MK}. }
\label{fig:variance_MK}
\end{figure}

\subsection{variance of RB}

In the old theory, the variance is \cite{wickramasinghe05}
\begin{equation}\label{eq:variance_RB}
\begin{split}
&\Sigma(\eps; E; \alpha)=
\sqrt{\fr{\eps}{\pi^5 \Delta}} \\
&\Bigg[ \sum_{M_1,M_2 = 1}
4\fr{\sin^2[E \sqrt{\fr{\pi\Delta}{\eps} (\ti{M_1}^2 + \ti{M_2}^2 )}]}{(\ti{M_1}^2 + \ti{M_2}^2 )^{3/2}} \\
&
+ \sum_{M_1 = 1} \fr{\sin^2[E\sqrt{ \fr{\pi\Delta}{\eps} \ti{M_1}^2 }]}{\ti{M_1}^3 }
+ \sum_{M_2 = 1} \fr{\sin^2[E\sqrt{ \fr{\pi\Delta}{\eps} \ti{M_2}^2 }]}{\ti{M_2}^3 } \Bigg].
\end{split}
\end{equation}

In the new theory, the variance is
\begin{equation}\label{eq:variance_coherent_RB}
\begin{split}
&\Sigma(\eps; E; \alpha) \\
&=
\sqrt{\fr{\eps}{\pi^5 \Delta}}
\bigg[ \sum_{M_1,M_2 = 1}
4\fr{\sin^2[ \fr{\pi\Delta}{\eps} (\ti{M_1}^2 + \ti{M_2}^2 )]}{(\ti{M_1}^2 + \ti{M_2}^2 )^{3/2}} \\
&
+ \sum_{M_1 = 1} \fr{1}{2}\lf(1 + \lf(1 - \sqrt{\ti{M_1}}(\fr{\pi}{\eps})^{1/4} \rg)^2 \rg)
\fr{\sin^2[ \fr{\pi\Delta}{\eps} \ti{M_1}^2 ]}{\ti{M_1}^3 } \\
&+ \sum_{M_2 = 1} \fr{1}{2}\lf(1 + \lf(1 - \sqrt{\ti{M_2}}(\fr{\pi}{\eps})^{1/4} \rg)^2 \rg)
\fr{\sin^2[ \fr{\pi\Delta}{\eps} \ti{M_2}^2 ]}{\ti{M_2}^3 } \bigg].
\end{split}
\end{equation}

Unlike the modified Kepler problem, the average of an ensemble of $\alpha$ does not converge to the average value of $\alpha$.
Hence the correct variance should be given by
\begin{equation}\label{eq:Sigma_ensemble_average}
\int\Sigma(\eps; E; \alpha) \rho(\alpha) d\alpha,
\end{equation}
which does not converge to $\Sigma(\eps; E; \alpha_{\mbox{average}})$.

\begin{figure}[htp]
\begin{center}
\includegraphics[width=3.0in]{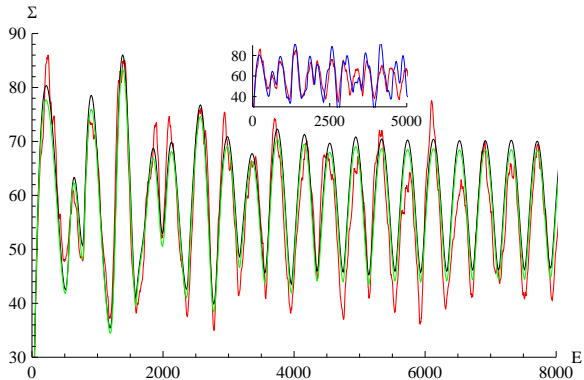}
\end{center}
\caption{\label{fig:variance_RB}
The red line is the numerical result.
The blue line is the theoretical result without ensemble averaging over $\alpha$.
The green line is theoretical fit using Eq. \ref{eq:variance_coherent_RB} with ensemble averaging.
The black line is theoretical fit using Eq. \ref{eq:variance_RB} with ensemble averaging.
We see that the ensemble averaging fits well in all the range, while it fits well only for $E\le 3000$ without ensemble averaging.
The variance with coherent interference is a little better than without. }
\end{figure}

\section{Level rigidity}

\subsection{rigidity as function of energy interval width}

\begin{figure}[htp]
\begin{center}
\includegraphics[width=3.0in]{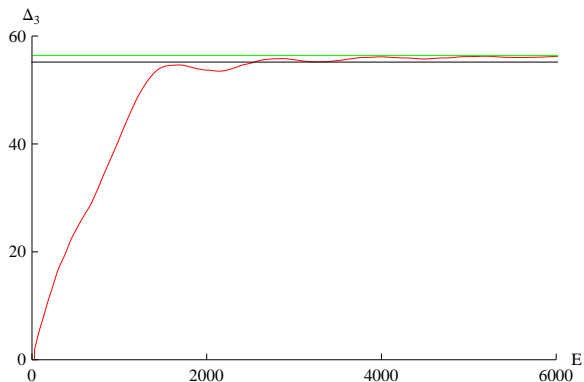}
\end{center}
\caption{\label{fig:RigidityEnergyWidth_MK} Rigidity of MK as a function of the energy interval width. $\eps = 2\times 10^5$. }
\end{figure}

\begin{figure}[htp]
\begin{center}
\includegraphics[width=3.0in]{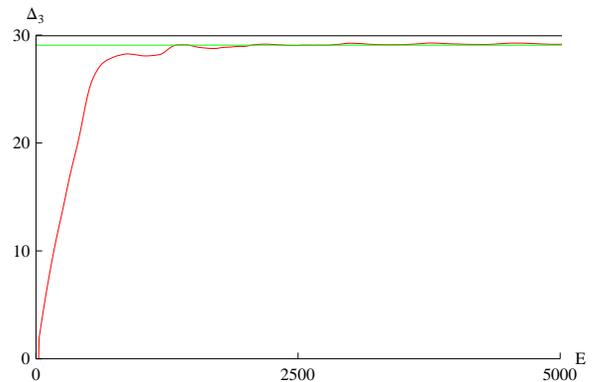}
\end{center}
\caption{\label{fig:RigidityEnergyWidth_RB} Rigidity of RB as a function of the energy interval width. $\eps = 10^5$. }
\end{figure}

\subsection{rigidity of MK}

The saturation level rigidity is calculated by
\begin{equation}
\Delta_{3, MK}(\eps, E)=\fr{2}{E^4}\int_0^{E} dx(\eps^3-2x \eps^2+x^3)\Sigma(\eps, x).
\end{equation}

\begin{equation}\label{eq:rigidity_MK}
\Delta_3(\eps, E)=
\sum_{M_r = 1}\lf(\lf\lfloor \fr{M_r}{\gamma^{(cir)}}\rg\rfloor + \fr{1}{4} \rg)
\fr{\om }{2\pi ^2 M_r^3}.
\end{equation}

Compared with the Eq. 69 of \cite{wickramasinghe08}, the extra contribution to from radial orbits
\begin{equation}
\fr{1}{4}\sum_{M_r>0} \fr{\om }{2\pi ^2 M_r^3}=\fr{\om}{2\pi^2} \zeta (3)=37.29 .
\end{equation}

\subsubsection{long term trend of rigidity}

For $M_r = 1$ and $(\fr{3\eps}{2\beta})^{1/3} \gg 1$,
\begin{equation}
\Delta_3^\infty
\approx \text{const} + \fr{(2\beta)^{1/6}}{12} (3\eps)^{1/3}.
\end{equation}
But for $\beta = 3\times 10^6$, this requires $\eps \gg 3\times 10^6$.

\subsubsection{discrepancy of rigidity}

\begin{figure}[htp]
\begin{center}
\includegraphics[width=3.0in]{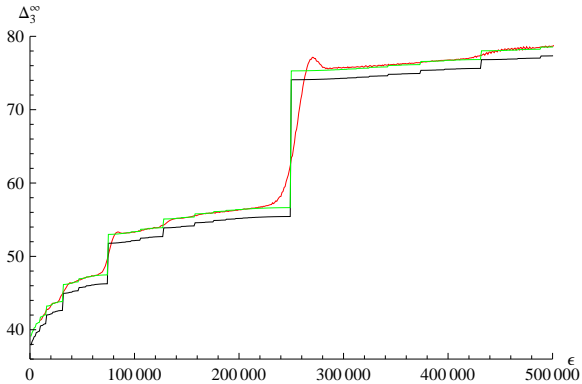}
\end{center}
\caption{We use the same values of $\beta$ as in Ref. \cite{wickramasinghe08}.
The red line is the numerical result.
The black line is the theoretical result from Eq. \ref{eq:rigidity_MK}.
The green line is the theoretical result from Eq. \ref{eq:rigidity_coherent_MK}. }\label{fig:rigidityTheoryNumericCompare_MK}
\end{figure}

In Fig. \ref{fig:rigidityTheoryNumericCompare_MK}, there is discrepancy between numerical and theoretical results of rigidity.
We explain it in this section.

The new rigidity is that
\begin{equation}\label{eq:rigidity_coherent_MK}
\begin{split}
&\Delta_3(\eps, E)=
\sum_{M_r = 1}\lf(\lf\lfloor \fr{M_r}{\gamma^{(cir)}}\rg\rfloor + \fr{1}{8}
+ \fr{1}{8} \lf(1 + \sqrt\fr{2M_r}{\om} \rg)^2 \rg) \\
&\fr{\om }{2\pi ^2 M_r^3}.
\end{split}
\end{equation}

\subsection{rigidity of RB}

In the old theory, the rigidity is
\begin{equation}\label{eq:rigidity_RB}
\begin{split}
\Delta_3 (\eps)
&= \fr{1}{\pi^{5/2}} \eps^{1/2} \Bigg[\sum_{M_1,M_2>0} \fr{1}{(\ti{M_1}^2 + \ti{M_2}^2)^{3/2}} \\
&+ \sum_{M_1>0}\fr{1/4}{\ti{M_1}^3} + \sum_{M_2>0}\fr{1/4}{\ti{M_2}^3} ].
\end{split}
\end{equation}

In the new theory, the rigidity is
\begin{equation}\label{eq:rigidity_coherent_RB}
\begin{split}
\Delta_3 (\eps)
&= \fr{1}{\pi^{5/2}} \eps^{1/2} [\sum_{M_1>0,M_2>0} \fr{1}{(\ti{M_1}^2 + \ti{M_2}^2)^{3/2}} \\
&+
\sum_{M_1>0}\fr{1/8}{\ti{M_1}^3} +
\sum_{M_1>0}\fr{1/8}{\ti{M_1}^3} (1 - \sqrt{\ti{M_1}} (\fr{\pi}{\eps})^{1/4})^2 ] \\
&+
\sum_{M_2>0}\fr{1/8}{\ti{M_2}^3} +
\sum_{M_2>0}\fr{1/8}{\ti{M_2}^3} (1 - \sqrt{\ti{M_2}} (\fr{\pi}{\eps})^{1/4})^2 ] .
\end{split}
\end{equation}

\begin{figure}[htp]
\begin{center}
\includegraphics[width=3.0in]{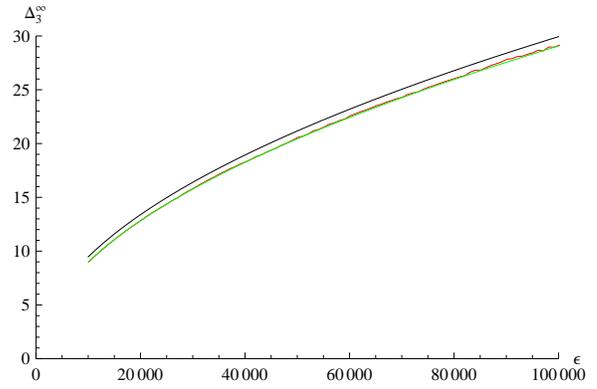}
\end{center}
\caption{The rigidity of RB.
The red line is the numerical result.
The black line the old theoretical result from Eq. \ref{eq:rigidity_RB}.
The green line the new theoretical result from Eq. \ref{eq:rigidity_coherent_RB}.
The inset shows the rigidity change against energy width and $\eps = 10^5$.
}
\end{figure}

\section{Conclusions}

We presented a complete quantum mechanical derivation of the level correlation function of a modified Kepler problem. With the inclusion of a class of periodic orbits omitted in \cite{wickramasinghe08} (as well as number of other corrections), it is in full agreement with the periodic orbit theory.

The account of long-range level correlations leads to striking properties of the level number variance and saturation level rigidity, as given by eqs. (\ref{eq:variance_coherent_MK}) and  (\ref{eq:rigidity_coherent_MK}). The latter are the main results of this work and are in excellent agreement with the numerical calculation, as shown in Figs. \ref{fig:variance_MK} and \ref{fig:rigidityTheoryNumericCompare_MK}. 

\appendix

\section{Derivation of correlation function, variance and rigidity}

\subsection{fluctuation of the level density}

The level density in the variable $\mathbf{k}$ has two parts $\rho(\mathbf{k}) = \rho^{(2)}(\mathbf{k}) + \rho^{(1)}(\mathbf{k})$, where $\mathbf{k}=(l,p)$, $\rho^{(2)}(\mathbf{k})$ is a delta function at lattice points in the first quadrant (``1Q'') and half delta function at lattice points on the half-axes, $\rho^{(1)}(\mathbf{k})$ is half delta function at lattice points on the half-axes.
The level density in the variable $\eps$ is \cite{wickramasinghe05}
\begin{equation}
\rho(\eps) \approx \int_{\text{1Q}} d^2 \mathbf{k} \delta(\eps-\phi(\mathbf{k})) \rho(\mathbf{k}),
\end{equation}
where $\phi(\mathbf{k})= (2p\om+ l^2)^{3/2} / 3\om$. To use the Poisson summation formula, we make $\rho(\mathbf{k})$ fully periodic by symmetrizing the integral to the whole plane (``WP''). Then
\begin{equation}
\rho(\eps) \approx\fr{1}{4} \int_{\text{WP}} d^2 \mathbf{k} \delta(\eps-\phi '(\mathbf{k})) \rho(\mathbf{k}),
\end{equation}
where $\phi'(\mathbf{k})=(2|p|\om+ l^2)^{3/2}/3\om$.

We consider $\rho^{(2)}(\eps)$ first. After applying Poisson summation formula, $\rho^{(2)}(\eps)$ becomes a summation over Fourier component $I^{(2)}_{\mathbf{w}}$ with $\mathbf{w} = 2\pi(M_\theta, M_r)$ with integer $M_\theta$ and $M_r$:
\begin{equation}
\rho^{(2)}(\eps)
= \sum_{M_\theta,M_r} I^{(2)}_{\mathbf{w}}(\eps).
\end{equation}
A Fourier component of $\rho^{(2)}(\eps)$ looks like
\begin{equation}
I^{(2)}_\mathbf{w}(\eps) =\fr{1}{4} \int_{L_\eps , WP} d k_{\parallel} \fr{1}{|\nabla_k \phi'(\mathbf{k})|} e^{i \mathbf{w} \cdot \mathbf{k}},
\end{equation}
where $L_\eps$ is the constant energy surface.
For the case of $M_\theta = M_r = 0$,
\begin{equation}
I^{(2)}_{2\pi(0,0)}(\eps)
= \int_{\text{1Q}} d k_{\parallel} \fr{1}{|\nabla_k \phi|}
= 1
\end{equation}
which is the average level density.
The $\mathbf{w}=0$ component gives the average level density and other components give the fluctuation $\delta\rho(\eps)$.
The fluctuation of level density is given by
\begin{equation}
\delta\rho^{(2)}(\eps)
=\lf( \sum_{M_\theta,M_r \neq 0} + \sum_{M_\theta = 0,M_r \neq 0} + \sum_{M_\theta \neq 0,M_r=0} \rg) I^{(2)}_{\mathbf{w}}(\eps).
\end{equation}

\subsubsection{case of $M_\theta, M_r>0$}

For every $\mathbf{w}=2\pi(M_\theta,M_r)$ with $M_\theta, M_r>0$, there are four symmetric $\mathbf{w}$s: $2\pi(\pm M_\theta,\pm M_r)$. We redefine $I^{(2)}_{\mathbf{w}}(\eps)$ as the sum of these four terms:
\begin{equation}\label{eq:sum4}
I^{(2)}_{\mathbf{w}}(\eps)
= \int_{\text{WP}} d k_{\parallel} \fr{1}{|\nabla_k \phi'|} \cos(2\pi M_\theta l)\cos(2\pi M_r p)
\end{equation}
where $M_\theta, M_r>0$.
Also notice that the integral is symmetric between positive and negative $l$ or $p$.
Using this symmetry, we return to the first quadrant.
\begin{equation}
\begin{split}
&I^{(2)}_{\mathbf{w}}(\eps)
= 4 \int_{\text{1Q}} d k_{\parallel} \fr{1}{|\nabla_k \phi|} \cos(2\pi M_\theta l)\cos(2\pi M_r p) \\
&
= \fr{1}{(3\om\eps)^{1/3}}\sqrt{\fr{2\om}{M_r}} \times \\
&\lf[ \cos \lf(\fr{\pi M_r^2 (3\om\eps)^{2/3}+\pi M_\theta^2\om^2}{M_r\om}\rg)(C_{-}(\eps)+C_{+}(\eps))+\rg.\\
&\lf. \hspace{6pt}
      \sin \lf(\fr{\pi M_r^2 (3\om\eps)^{2/3}+\pi M_\theta^2\om^2}{M_r\om}\rg)(S_{-}(\eps)+S_{+}(\eps)) \rg],
\end{split}
\end{equation}
where $\text{C/S}_{\pm}(\eps)=\text{C/S}(\sqrt{\fr{2}{M_r\om}}(M_r(3\om\eps)^{1/3}\pm M_\theta\om))$ and $C(x)$ and $S(x)$ are the Fresnel integrals defined as $^C_S(x)=\int_0^x dt\;^{\cos}_{\sin}(\fr{\pi}{2}t^2)$.
The Fresnel integrals are almost step function. We treat $C_{+}(\eps) = S_{+}(\eps) = \fr{1}{2}$, if $\fr{M_\theta\om}{\sqrt{M_r \om}}\gg 1$, which is true for large $\beta$.

The terms $C_{-}(\eps)$ and $S_{-}(\eps)$ are important because they are almost step functions and they are the underlying origin of quantum jumps. If
\begin{equation}
\sqrt{\fr{2}{M_r\om}}(M_r(3\om\eps)^{1/3}- M_\theta\om)<1,
\end{equation}
both $C_{-}(\eps)$ and $S_{-}(\eps)$ are $-\fr{1}{2}$, both $C_{-}(\eps)+C_{+}(\eps)$ and $S_{-}(\eps)+S_{+}(\eps)$ are 0, which makes the oscillation terms
\begin{equation}
^{\cos}_{\sin} \lf(\fr{\pi M_r^2 (3\om\eps)^{2/3}+\pi n^2\om^2}{M_r\om}\rg)
\end{equation}
not important.

To simplify the problem, we approximate $C_{-}(\eps)+C_{+}(\eps)$ and $S_{-}(\eps)+S_{+}(\eps)$ using the step function.
\begin{equation}
C/S_{-}(\eps)+C/S_{+}(\eps)\approx \begin{cases}
1 & \text{if\,} M_\theta \leq \fr{M_r}{\gamma^{(cir)}}, \\
0 & \text{otherwise} .
\end{cases}
\end{equation}

Then $I^{(2)}_{2\pi(M_\theta,M_r)}(\eps)$ can be approximated as
\begin{equation}\label{eq:InmSimple}
\begin{split}
& I^{(2)}_{2\pi(M_\theta,M_r)}(\eps)
\approx
\sum_{M_\theta \leq \fr{M_r}{\gamma^{(cir)}}}
\fr{2}{(3\om\eps)^{1/3}}\sqrt{\fr{\om}{M_r}} \\
&\cos \lf(\fr{\pi M_r^2 (3\om\eps)^{2/3}+\pi M_\theta^2\om^2}{M_r\om}-\fr{\pi}{4}\rg).
\end{split}
\end{equation}

\paragraph{Quantum jumps}

The quantum jumps happen when
\begin{equation}
M_\theta = \fr{M_r}{\gamma^{(cir)}}.
\end{equation}
For $\beta=3\times 10^6$, when $M_\theta=1$ and $M_r=2$, $\eps=2.5\times 10^5$.
When $M_\theta=1$ and $M_r=1$, another big jump will happen at $\eps=2\times 10^6$, which is not shown in Fig. 1 of \cite{wickramasinghe08}.
When $M_\theta=1$ and $M_r=3$, $\eps=0.74\times 10^5$. This jump is also in Fig. \ref{fig:rigidityTheoryNumericCompare_MK}.

\paragraph{Quantum and classical correspondence}

$M_r$ and $M_\theta$ correspond to winding numbers of the classical orbits in the angular direction and radial direction respectively.
The orbit periods are always multiples of $T_{min} = \fr{2\pi}{\om}$. When $M_\theta=1$ and $M_r=1$ and $\eps=2\times 10^6$, the system has reached its minimal period.

\subsubsection{case of $M_\theta=0$, $M_r>0$}

There are two symmetric terms in Eq. \ref{eq:sum4}.
\begin{equation}\label{eq:I0Mr}
\begin{split}
&I^{(2)}_{2\pi(0,M_r)}(\eps)\\
&= 2 \int_{\text{1Q}} d k_{\parallel} \fr{1}{|\nabla_k \phi|} \cos(2\pi M_r p)\\
&\approx \fr{1}{(3\om\eps)^{1/3}} \sqrt{\fr{\om}{M_r}}
\cos(\fr{\pi M_r (3\om\eps)^{2/3}}{\om}-\fr{\pi}{4}).
\end{split}
\end{equation}
Classically, $I^{(2)}_{2\pi(0,M_r)}(\eps)$ corresponds to radial periodic orbits.
The rigidity before jumps comes from radial periodic orbits.

\subsubsection{case of $M_\theta>0$ and $M_r=0$}

There are two symmetric terms in Eq. \ref{eq:sum4}.
\begin{equation}\label{eq:IMtheta0}
\begin{split}
I^{(2)}_{2\pi(M_\theta,0)}(\eps)
&= 2 \int_{\text{1Q}} d k_{\parallel} \fr{1}{|\nabla_k \phi|} \cos(2\pi M_\theta l) \\
&= \fr{1}{(3\om\eps)^{1/3}}\fr{1}{\pi M_\theta}\sin(2\pi M_\theta (3\om\eps)^{1/3}).
\end{split}
\end{equation}
Classically, $I^{(2)}_{2\pi(M_\theta,0)}(\eps)$ corresponds to isolated angular periodic orbits
After collecting Eq. \ref{eq:InmSimple}, \ref{eq:I0Mr}, \ref{eq:IMtheta0}, we get Eq. \ref{eq:deltaRhoEpsilon}.

\paragraph{stable isolated periodic orbits}

Comparing $I^{(2)}_{2\pi(M_\theta,0)}(\eps)$ with $I^{(2)}_{2\pi(M_\theta,M_r)}(\eps)$,
\begin{equation}
\fr{I^{(2)}_{2\pi(M_\theta,0)}(\eps)}{I^{(2)}_{2\pi(M_\theta,M_r)}(\eps)}
\sim
\fr{\sqrt{M_r}}{2\pi M_\theta}\fr{1}{\sqrt{\om}}
\sim
\fr{1}{2\pi \sqrt{\om}}=0.0032.
\end{equation}
$I^{(2)}_{2\pi(M_\theta,0)}(\eps)$ is far smaller than other terms.
The saturation rigidity due to these stable isolated periodic orbits is
\begin{equation}
\Delta_3(\eps)=
\sum_{M_\theta>0}
\fr{(3\om\eps)^{2/3}}{8\pi^4 M_\theta^4 \om^2} ,
\end{equation}
which is smaller than $6\times 10^{-4}$ at $\eps = 5\times 10^5$.
The reason is that the periodic orbit of $M_r\neq 0$ actually denotes a family of continuous periodic orbits with the same $M_\theta$ and $M_r$. The superposition of a family of periodic orbits contribute  the fluctuation of levels far more significantly than stable isolated periodic orbits. \footnote{From the perspective of particle-wave duality, a single periodic orbit can hardly support a quantum state. From the perspective of Thomas-Fermi approximation, the contribution to the density of state from the small phase space volume of isolated periodic orbits must be small.}

\subsection{coherent interference}

Now we calculate the Balian-Bloch term $\rho^{(1)}(\mathbf{k})$.
Because the semiclassical result is calculated from the assumption of fully periodic $\rho(k)$, the contribution from $\delta\rho^{(1)}(\eps)$ is non-semiclassical \cite{wickramasinghe05}.

On the $l$-axis,
\begin{equation}
\delta\rho^{(1)}_{l-axis}(\eps)
=\sum_{\nu> 0} \fr{\om}{(3\om\eps)^{2/3}}\cos(w_\nu (3\om\eps)^{1/3}),
\end{equation}
where $w_\nu=2\pi\nu$.
\begin{equation}
\Delta_3(\eps)
=\sum_{\nu>0}\fr{1}{8\nu^2 \pi^2}
=\fr{1}{48}.
\end{equation}
We can ignore this contribution to the final variance and rigidity.
On the $p$-axis,
\begin{equation}
\delta\rho^{(1)}_{p-axis}(\eps)
=\sum_{\nu>0}\fr{1}{(3\om\eps)^{1/3}}\cos(w_\nu \fr{(3\om\eps)^{2/3}}{2\om}).
\end{equation}
$\Delta_3(\eps) = \fr{1}{48}$.

Although $\delta\rho^{(1)}_{l-axis}(\eps)$ is coherent with $I^{(2)}_{2\pi(M_\theta,0)}(\eps)$, the coherent contribution to rigidity is small because $I^{(2)}_{2\pi(M_\theta,0)}(\eps)$ is very small.
We consider the coherent interference between $\delta\rho^{(1)}_{p-axis}(\eps)$ and $I^{(2)}_{2\pi(0, M_r)}(\eps)$.
\begin{equation}
\begin{split}
&\delta\rho^{(1)}_{p-axis}(\eps)+\sum_{M_r}I^{(2)}_{2\pi(0, M_r)}(\eps)\\
& =\sum_{\nu>0}\fr{1}{(3\om\eps)^{1/3}} \sqrt{\fr{\om}{2\nu}}
\sin(\fr{\pi \nu (3\om\eps)^{2/3}}{\om})\\
&+\sum_{\nu>0}\fr{1}{(3\om\eps)^{1/3}}\lf( 1+\sqrt{\fr{\om}{2\nu}} \rg)\cos(\pi\nu \fr{(3\om\eps)^{2/3}}{\om})
\end{split}
\end{equation}
The $\sin$ term still has no interference. Its contribution to rigidity is the same as the previous theory.
Comparing with $\delta\rho^{(2)}(\eps)$, the only difference is that we have an extra factor $1+\sqrt{\fr{\om}{2\nu}}$.
In the same way, in variance and rigidity, we need to add an extra factor like $\lf(1+\sqrt{\fr{\om}{2\nu}}\rg)^2$ before corresponding terms.

\subsection{calculation of RB}

A similar calculation gives
\begin{equation}
I^{(2)}_{\vec \nu}=4J_0(2\pi \nu k) = 4J_0(2\pi \sqrt{ \ti{M_1}^2 + \ti{M_2}^2} k);
\end{equation}
for one of $M_1$ and $M_2$ equal to zero,
$ I^{(2)}_{\vec \nu}=2J_0(2\pi \ti{M_1} k_x)$, or $ I^{(2)}_{\vec \nu}=2J_0(2\pi \ti{M_2} k_y)$.
$I^{(2)}_{\vec \nu}=J_0(0) = 1$,
which is the average spectral staircase.
The two parts of $\rho^{(1)}$ is
\begin{equation}
-\sum_{M>0}
\fr{1}{ \sqrt{\pi\eps }}\cos(2\pi \ti{M} \sqrt{\fr{4\eps}{\pi}}).
\end{equation}
for $M = M_1$ or $M_2$.

\subsubsection{coherent interference}

Bessel function $J_0$ has the asymptotic form
\begin{equation}
J_0(x) \approx \sqrt\fr{2}{\pi x}\cos(x-\fr{\pi}{4}).
\end{equation}
The only coherent interference term comes from
\begin{equation}
\begin{split}
&2J_0 \lf(2\pi \ti{M} \sqrt{\fr{4\eps}{\pi}} \rg) - \fr{1}{ \sqrt{\pi\eps }}\cos \lf(2\pi \ti{M} \sqrt{\fr{4\eps}{\pi}} \rg) \\
&= \sum_{M>0}
\fr{1}{\pi\sqrt{ \ti{M} \sqrt{\fr{\eps}{\pi}} }} \Bigg[
\sin\lf(2\pi \ti{M} \sqrt{\fr{4\eps}{\pi}} \rg) \\
&+\lf( 1 - \sqrt{\ti{M}} (\fr{\pi}{\eps})^{1/4} \rg)
\cos\lf(2\pi \ti{M} \sqrt{\fr{4\eps}{\pi}} \rg) \Bigg]
\end{split}
\end{equation}
for $\ti{M} = \ti{M_1}$ or $\ti{M_2}$.
For high energy, the coherent effect between $\rho^{(2)}$ and $\rho^{(1)}$ is suppressed by a factor $\fr{1}{\eps^{1/4}}$.
The above factor is smaller than 1. This means that the coherent interference decreases the rigidity and variance.

\section{Ensemble averaging}

\subsection{parametric averaging}

The ensemble average of a physical quantity can be mathematically defined as an integral over the distribution function of $\beta$ or $\om$. The ensemble average of $x$ is
\begin{equation}
\la x\ra = \int_{-\infty}^\infty x(\om) \rho(\om) d\om
\end{equation}
We calculate the ensemble average by sampling $\om$ from a normal distribution.

\subsection{average spectral staircase}

After ensemble averaging,
\begin{equation}
\la \delta\rho(\eps) \ra =0.
\end{equation}
This requires that
\begin{equation}\label{eq:average_oscillation_zero}
\begin{split}
&\la \cos (\fr{\pi M_r^2 (3\om\eps)^{2/3}+\pi M_\theta^2\om^2}{M_r\om}-\fr{\pi}{4}) \ra = 0 \\
&\la \cos(\fr{\pi M_r (3\om\eps)^{2/3}}{\om}-\fr{\pi}{4}) \ra = 0 \\
&\la \sin(2\pi M_\theta (3\om\eps)^{1/3}) \ra = 0.
\end{split}
\end{equation}
From another perspective, $\cdot$ in $\cos(\cdot)$ can be linearized to be a linear function of $\om$. The integration of $\cos(\cdot)$ over $\om$ gives zero.
The ensemble averaging in numerical calculation corresponds to the diagonal approximation in theoretical calculation.

\subsection{diagonal approximation}

Here we give a new and general argument of diagonal approximation. Though our argument is specific to MK, it seems to be applicable to other problems too.
There are many interference terms in correlation function of levels, such as $I^{(2)}_{2\pi(M_{\theta_1},M_{r_1})}(\eps_1)I^{(2)}_{2\pi(M_{\theta_2},M_{r_2})}(\eps_2)$ for $M_{\theta_1}\neq M_{\theta_2}$ or $M_{r_1}\neq M_{r_2}$.
The interference between $M_{\theta_1}\neq M_{\theta_2}$ or $M_{r_1}\neq M_{r_2}$
is
\begin{widetext}
\begin{equation}
\begin{split}
&I^{(2)}_{2\pi(M_{\theta_1},M_{r_1})}(\eps_1)I^{(2)}_{2\pi(M_{\theta_2},M_{r_2})}(\eps_2)\\
&\sim
\cos\lf[ \pi \fr{ M_{r_1}M_{r_2} \lf( M_{r_1}(3\eps_1)^{2/3}+M_{r_2}(3\eps_2)^{2/3} \rg)\om^{2/3} +
\lf(M_{\theta_1}^2 M_{r_2}+ M_{\theta_2}^2 M_{r_1} \rg) \om^2 }{M_{r_1}M_{r_2}\om}  -\fr{\pi}{2} \rg] + \\
&\hspace{12pt} \cos\lf[ \pi \fr{M_{r_1}M_{r_2} \lf( M_{r_1}(3\eps_1)^{2/3}-M_{r_2}(3\eps_2)^{2/3} \rg)\om^{2/3} +
\lf(M_{\theta_1}^2 M_{r_2}- M_{\theta_2}^2 M_{r_1} \rg) \om^2 }{M_{r_1}M_{r_2}\om} \rg]
\end{split}.
\end{equation}
\end{widetext}

Compare with Eq. \ref{eq:average_oscillation_zero}, the two terms of $\cos$ of the above equation have the same form. The first term is always zero after ensemble averaging. For the second term, if $M_{r_{1}}\neq M_{r_{2}}$ and $M_{\theta_1}\neq M_{\theta_2}$, it is also zero. This implies that the interference between two terms with $M_{r_{1}}\neq M_{r_{2}}$ and $M_{\theta_1}\neq M_{\theta_2}$ is zero after ensemble averaging.
If $M_{r_{1}} \neq M_{r_{2}}$ and $M_{\theta_1} = M_{\theta_2}$, the form of the second term is also similar to Eq. \ref{eq:average_oscillation_zero}. So it must be zero after ensemble averaging.
If $M_{r_{1}} = M_{r_{2}}$ and $M_{\theta_1}\neq M_{\theta_2}$, for $\eps_1 \neq \eps_2$, the second term is zero after ensemble averaging. But it is indeed somewhat different because here the factor
\begin{equation}
\eps_1^{2/3} - \eps_2^{2/3},
\end{equation}
is small. This may makes Eq. \ref{eq:average_oscillation_zero} not applicable any more.  But the main part of the second term is that
\begin{equation}
\fr{ (\pi M_{\theta_1}^2 M_{r_2}-\pi M_{\theta_2}^2 M_{r_1}) \om^2 }{M_{r_1}M_{r_2}\om}  =
\fr{ (\pi M_{\theta_1}^2 M_{r_2}-\pi M_{\theta_2}^2 M_{r_1}) \om }{M_{r_1}M_{r_2}}
\end{equation}
quickly oscillates with $\om$ and becomes zero after ensemble averaging.

In summary, the interference terms between different periodic orbits with $M_{r_{1}} \neq M_{r_{2}}$ or $M_{\theta_1} \neq M_{\theta_2}$ must be zero. The ensemble averaging removes the interference between different orbits and makes the correlation function, variance and rigidity become a sum over classical orbits. In this way, we give a ``proof'' diagonal approximation of MK from the perspective of ensemble averaging.

\subsection{diagonal approximation for RB}

A typical non-diagonal term in the level correction function looks like
\begin{equation}
\begin{split}
&
\la \cos(2\pi \ti{M_1} \sqrt{\fr{4\eps_1}{\pi}}-\fr{\pi}{4})
\cos(2\pi \ti{M_2} \sqrt{\fr{4\eps_2}{\pi}}-\fr{\pi}{4}) \ra \\
&\sim \la
\cos( 2\pi \ti{M_1} \sqrt{\fr{4\eps_1}{\pi}} - 2\pi \ti{M_2} \sqrt{\fr{4\eps_2}{\pi}} ) \ra + \\
&\hspace{15pt} \la \sin( 2\pi \ti{M_1} \sqrt{\fr{4\eps_1}{\pi}} + 2\pi \ti{M_2} \sqrt{\fr{4\eps_2}{\pi}} )\ra .
\end{split}
\end{equation}
After replacing $x = \alpha^{1/4}$, the derivative of $x$ in $\cos(\cdot)$ is
$\sim M_1 \sqrt{\fr{4\eps_1}{\pi}} + \fr{1}{x^2} M_2 \sqrt{\fr{4\eps_2}{\pi}} \gg 0$. The integration over $\alpha$ gives 0.
It works in the same way for the $\sin$ term with $M_1 \neq M_2$.

When $M_1 = M_2$, the ensemble averaging over $\alpha$ does not give 0.
It is not similar to a term in $\delta\rho^{(2)}$ as its derivative of $x$ vanishes.
This term gives 0 for rigidity. But its contribution to correlation function and variance is not zero.
This needs further investigation. 

\section{Scaling of variance of MK}

\begin{figure}[htp]
\begin{center}
\includegraphics[width=3.2in]{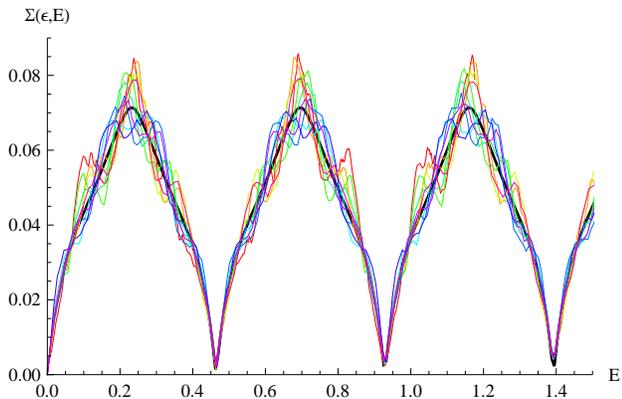}
\end{center}
\caption{Variance scaled. $\beta$ is $\beta = 1, 3, 5, \cdots, 19 \times 10^6$. The black line is the theoretical result without considering $\rho^{(1)}$. }
\end{figure}

\end{document}